\documentclass[useAMS,usenatbib]{mn2e}
\usepackage{epsfig,amssymb,amsmath}

\title[Transient nature of J195754+353513]{Transient nature of J195754+353513}
\author[Pal et al.]{Sabyasachi Pal$^{1}$\thanks{E-mail: sabya@csp.res.in (SP)} Dusmanta Patra$^{1}$ Monique Hollick$^{2, 3}$ Sandip K. Chakrabarti$^{4, 1}$\\
$^{1}$Indian Centre for Space Physics, 43 Chalantika, Garia Station Road, 700084, India\\
$^{2}$Defense Science and Technology Group, West Avenue, Edinburgh, South Australia, 5111, Australia\\
$^{3}$University of New South Wales, Anzac Parade, Kensington, NSW, 2052, Australia\\
$^{4}$S. N. Bose National Centre for Basic Sciences, Block-JD, Sector-III, Salt Lake, Kolkata, 700098, India}

\begin{document}

\date{}

\pagerange{\pageref{firstpage}--\pageref{lastpage}} \pubyear{2017}

\maketitle

\label{firstpage}

\begin{abstract}
We have searched for transient and/or variable radio sources in the field of Galactic micro-quasar Cygnus X-1 in 1.4 GHz (L band) using data from the Karl G. Jansky Very Large Array (JVLA). We used twenty years of data between 1983 and 2003. We found a source J195754+353513 showing transient behavior. The source was also mentioned earlier in NVSS and WENSS catalog but its transient nature was not reported earlier. The source is located 23.8 arcminutes far from Cygnus X-1. It is detected many times during the span of our study and it varied between less than 0.3 mJy to 201 mJy. J195754+353513 also showed high intra-day variability. In one occasion, the source rose from $\sim20$ mJy to $\sim180$ mJy 700 second. For limited number of cases circularly polarized emission could be detected from the source with $V/I$ vary between 0.15 to 0.25. 2MASS J19575420+3535152 may be the near-infrared counterpart of the source. We compared the properties of the source with other Galactic transient sources having similar properties. The nature of the source is still unknown. We speculate on its possible nature.

\end{abstract}


\section{Introduction} 

The steady state radio sky is considerably well studied and modeled (e.g. \citet{Be95, Co98}) but dynamic radio sky is not studied in detail due to one or more reasons including lack of telescopes with large field of view, observational 
constraints, band width and poor time resolution etc. There are various types of sources which show variability 
at different time scales. For example, brown dwarfs, flaring stars, masers, pulsars, micro-quasars, supernovae, gamma-ray bursts and active galactic nuclei show high levels of variability (e.g. \citet{Co04} for a review). Until recently there are only few radio surveys  
which searched for variable and transient radio sources efficiently (e.g., see, \citep{Gr86, Ma09}). So, most of the radio transients have been found through follow-up observations of known or suspected 
transient emitters. 

Recent programs to search for radio transients from direct and archival observations revealed some potential radio transients which are consistent with the expectation that previous limitations on the detection 
of radio transients were instrumental and not astrophysical. For example, A giant burst was detected from a young stellar object \citep{Bo03}. Many 1--3 Jy radio bursts were found at high and low Galactic latitudes \citep{Ni07, Ma07, Ki08}. A periodic, coherent and circularly polarized burst was found in an ultra-cool dwarf \citep{Ha07}.  
A few tens of Fast Radio Bursts (FRBs) have been detected so far which lasted only for a few milliseconds (e.g. \citet{Lo07, Ra16}). The origin of FRBs is still unclear but due to relatively high dispersion measures, some believe these sources have extra-galactic origin (e.g. \citet{Lo07}).
 
Huge improvements in field of view, especially at low radio frequencies, helps to study transient  and variable radio sources more effectively (e.g. \citet{Ma14}). However, no detection of any transient sources from recent $12000$ deg$^2$ systemic transient search comprising of 2800 pointings using with the Jansky Very Large Array Low-band Ionosphere and Transient Experiment (VLITE) \citep{Po16} and detection of only one transient source from monitoring of region close to the North Celestial Pole \citep{St16} covering 1925 deg$^2$ using LOFAR and no detection of transient source from 1430 deg$^2$ search using Murchison Wide-field Array (MWA) (\cite{Be14}; also see the MWA study by \citet{Ro16}) show that detections of transient radio sources are currently not very common, especially at low radio frequencies. Future deep, wide-field searches may potentially be far more fruitful (e.g. recent transient rate calculations in \citet{Me15, Mo16}).

Large archival data from various telescopes are important resource to look for transient and variable radio sources. Earlier, \citet{Ba11} reported 15 transient sources from the study of 22 yr archival data of the Molonglo Observatory Synthesis Telescope covering 2776 deg$^2$ survey area. Recently, \citet{Mu17} found a candidate transient source at low frequency from comparison of TIFR GMRT Sky Survey Alternative Data Release 1 (TGSS ADR1, see \citet{In17}) and the GaLactic and Extragalactic All-sky Murchison Widefield Array (GLEAM, see \citet{Hu17}) survey catalogues. Many variable and transient radio sources were reported from archival data search of the NRAO VLA Sky Survey (NVSS) \citep{Co98} and FIRST survey \citep{Le02, Th11}.
Ten milli Jansky level transients were detected from 22 years of archival the Karl G. Jansky Very Large Array (JVLA) observations of a single field-of-view at 5 and 8.4 GHz \citep{Bo07}. Though it was shown later that more than half of the transient sources reported in \citet{Bo07} were due to data artifacts and rest of the sources had low signal-to-noise ratio $(S/N)$ than \citet{Bo07} to conclusively find transient nature of these sources \citep{Fr12}. 

Earlier detection of two transient radio sources GCRT J1745--3009 \citep{Hy05,Hy07,Ro10} and GCRT J1742--3001 \citep{Hy09} were made from systematic search near Galactic Center region using The Karl G. Jansky Very Large Array (JVLA) and Giant Meterwave Radio Telescope (GMRT). GCRT J1745--3009 is a unique source which was detected only three times in 2002 \citep{Hy05}, 2003 and 2004 \citep{Hy07}. In 2002, this source exhibited $\sim$10 minute long, $\sim$1 Jy peaked bursts with a $\sim$77 minute period. The emission from the source was coherent \citep{Hy05} with extremely steep spectral index ($\alpha=-13.5\pm3.0$) and high circular polarization \citep{Ro10}. All the three detections of the source were in 330 MHz. The characteristics of GCRT J1742--3001 did not match with any known mechanisms of emission in transient compact sources. As a result, the source seems to represent a member of a new class of coherently emitting objects. GCRT J1742--3001 was active for a few months with $\sim$150 mJy flux density at the peak and showed no periodicity in emission \citep{Hy09}. The source was detected in 235 MHz and exhibited steep spectral index ($\alpha<-2$). For both of these sources, no counterpart was discovered in high energy, making them impossible 
to be detected by conventional follow-up radio observations from high energy observations. 

In this paper, we look for new variable/transient sources from a single well observed field centered at micro-quasar Cygnus X-1. Though we could not
detect a new source, we discovered hitherto un-reported transient 
behavior of one NVSS source, namely, NVSS J195754+353513. This source is located approximately 
$23.8$ arcminutes far from the micro-quasar Cygnus X-1 at J2000 co-ordinates 19h57m54.3s 
($\pm$ 0.7s) +35$^{\circ}$34$^{\prime}$59.6$^{\prime \prime}$ ($\pm$0.6$^{\prime \prime}$). 
In Section 2, we summarize observational details and data analysis procedure. In Section 3,
we summarize various results on the source. We discuss significance of various 
findings in Section 4. Finally, we make concluding remarks in Section 5.

\begin{figure*}
\vbox{
\centerline{
\psfig{figure=B_07-11-1999-VS2.PS,height=7cm,width=7cm,angle=0}
\psfig{figure=B_11-02-2000-VS2.PS,height=7cm,width=7cm,angle=0}}}
\vbox{
\centerline{
\psfig{figure=A_04-03-90-VS2.PS,height=7cm,width=7cm,angle=0}
\psfig{figure=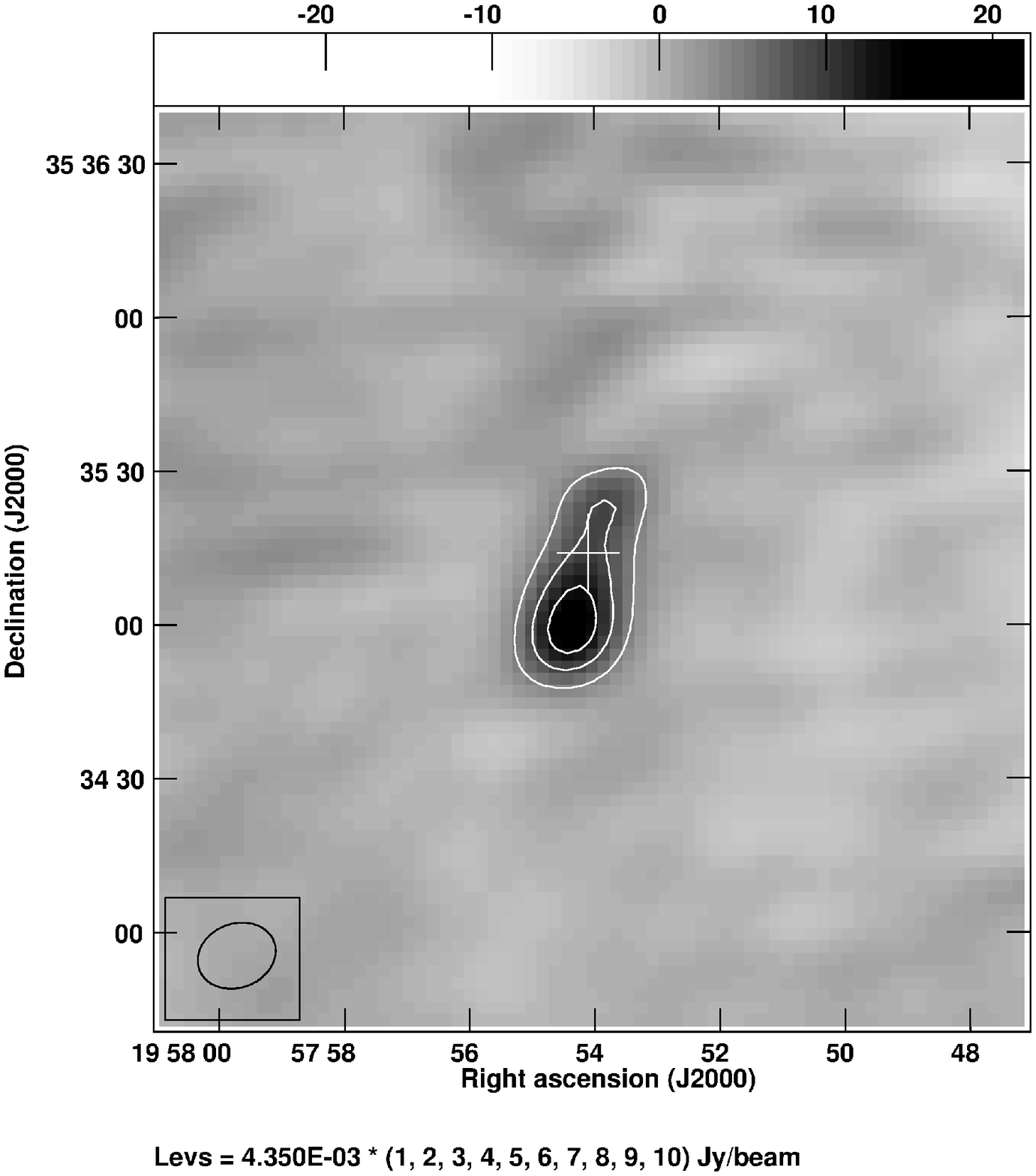,height=7cm,width=7cm,angle=0}}}
\vbox{
\centerline{
\psfig{figure=D_18-03-91-1-EDITED.PS,height=7cm,width=7cm,angle=0}
\psfig{figure=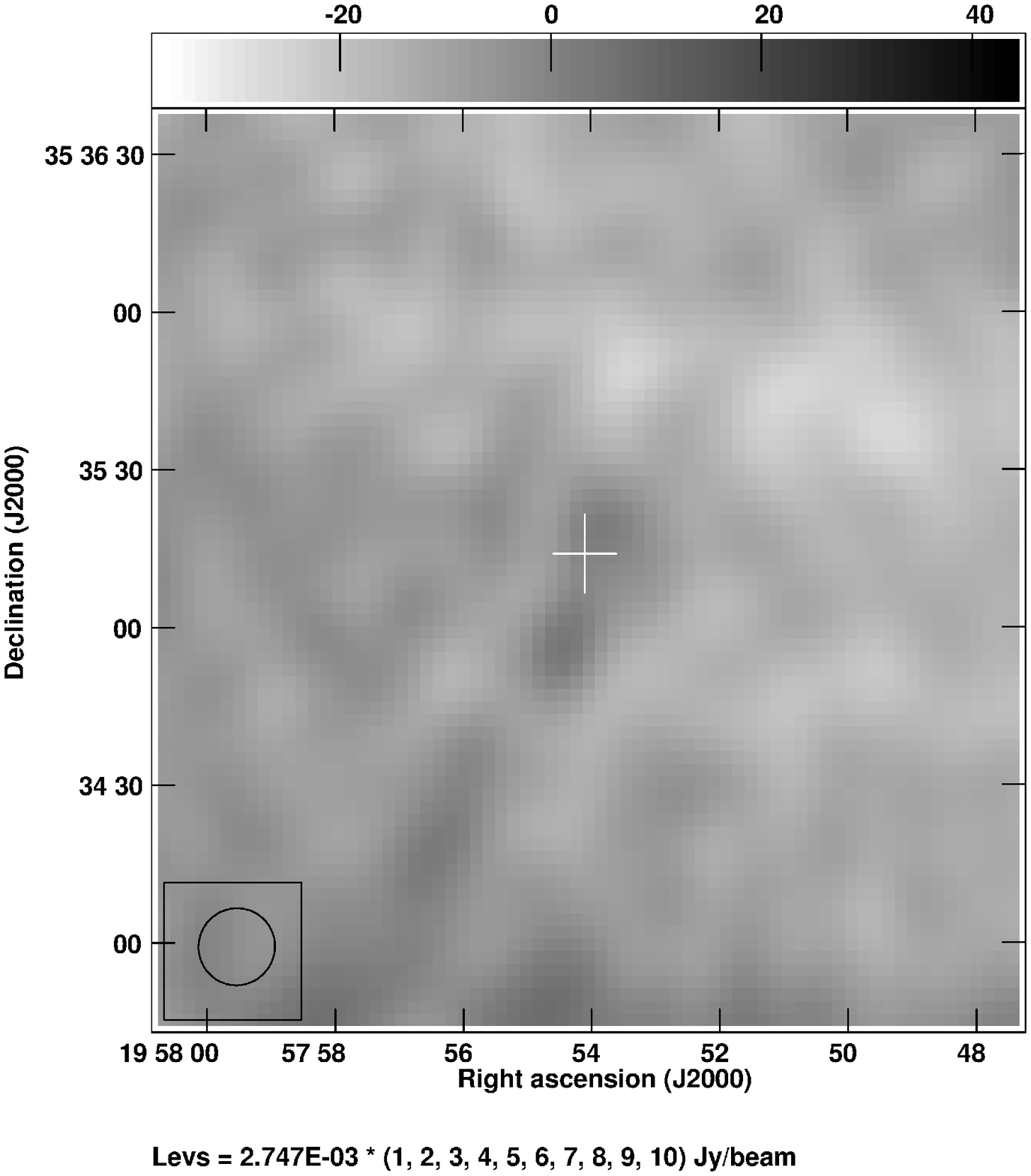,height=7cm,width=7cm,angle=0}}}
\caption{Images of J195754+353513 in different configurations of JVLA at 1490 MHz. In the upper left and right panels, we have shown images of the source in B configuration on 7th November 1999 and 11th February 2000 with flux density 56.5 mJy and 26.0 mJy respectively. In the middle left, middle right and lower left panels, we have shown examples of detection of the source in A, C and D configurations of JVLA. In the lower right panel, we have shown an example of no detection when JVLA was in C configuration. We have included the synthesized beam at the left-bottom corner of each panel. The location of the NVSS position is shown by a cross in all the images where size of the cross is error in NVSS location multiplied by 10 for easy visualization. For details, see text and Table 1.}
\label{transient-image}
\end{figure*}

\section{Observation and Data analysis with The Karl G. Jansky Very Large Array}

A blind search for new variable radio sources was conducted using archival 
JVLA data\footnote{https://science.nrao.edu/facilities/vla/archive/index} at the L-band frequency of 1400 MHz. 
L band is good for this kind of search with JVLA as it provides right balance between field of view and sensitivity. Though the field of view in 74 and 325 MHz band with JVLA will be higher, these bands have relatively poor sensitivity. Availability of archival data is also less in frequencies less than 1400 MHz. Also, for some sub-classes, the transient detection rate is higher at 1400 MHz, as discussed in Section 1.

JVLA comprises of twenty seven fully steerable antennas each with 25-meter diameter in a Y-shaped array. It is located around eighty km west of Socorro, New Mexico. Antennas are periodically moved in different configuration to achieve different scientific goals where the most expanded antenna configuration is known as A configuration and the most compact one is D configuration. B and C configurations are intermediate between A and D. Occasionally antennas are placed in a hybrid configurations, like AB or BC when some of the antennas are in one configuration and some of them in other. The maximum size of the baselines \(B_{max}\) in A, B, C and D configurations are 36.4, 11.1, 3.4 and 1.03 km respectively which corresponds to synthesized beam-width \((\theta_{HPBW})\) 1.4, 4.6, 15 and 49 arcsec respectively at 1400 MHz.   

The data used for the present work was taken from different configurations of the JVLA between 12 October 1983 (MJD 45619) and 4 June 2003 (MJD 52794). In total, we have used 262 different epochs of observations with various intervals ranging from days to years between successive observations. The antennas were in configuration A, AB, AD, B, BC, C, CD, and D for 53, 16, 5, 46, 15, 45, 26, and 56 days, respectively.

\begin{table*}
\caption{\bf Details of observations corresponding to Figure \ref{transient-image}}
\centering
\begin{tabular}{l c c c c c c c}
\hline
Image   &Date  &        Date          & Configuration  & Flux Density & RMS        &Synthesized beam&On source time\\
Location&(UT)  &       (MJD)          &                & (mJy)        & (mJy beam$^{-1}$) &                & (sec)        \\
\hline
Upper Left  &07/11/1999&51489& B &56.5&0.7&4.60$^{\prime \prime}\times$ 4.22$^{\prime \prime}$&170 \\
Upper Right &11/02/2000&51585& B &26.0&0.4&4.44$^{\prime \prime}\times$ 4.14$^{\prime \prime}$&120\\
Middle Left &04/03/1990&47954& A &56.2&0.3&1.28$^{\prime \prime}\times$ 1.17$^{\prime \prime}$&13820\\
Middle Right &06/05/1984&45826& C &53.0&1.6&15.56$^{\prime \prime}\times$ 12.48$^{\prime \prime}$&390\\
Lower Left  &18/03/1991&48333& D &80.7&2.0&19.72$^{\prime \prime}\times$ 19.72$^{\prime \prime}$&1140\\
Lower Right &31/08/1997&50691& C &--- &5.5&14.70$^{\prime \prime}\times$ 14.48$^{\prime \prime}$&100\\
\hline
\end{tabular}
\end{table*}

We studied a field centered on Cygnus X-1, a radio emitting X-ray binary \citep{Bo65, Gi72} with co-ordinates 19h58m21.676s +35$^{\circ}$12$^{'}$05.78$^{''}$ (J2000). This area of the sky was chosen because the field of Cygnus X-1 is one of the best studied Galactic black hole binaries using JVLA due to the fact that it is the first system widely accepted to contain a black hole \citep{Mu71, Ta72, Gi86}. This make the field ideal for looking variable/transient sources.

There are big data gap between April 1986 to April 1988, May 1991 to October 1996 as well as February 2001 to June 2003. The `on source' observation time in each epoch was between 2 to 15 minutes. Observing bandwidth for most of the days was 50 MHz.  

Analysis and imaging of the data was carried out with Astronomical Image Processing System (AIPS)\footnote{http://www.aips.nrao.edu}. Bad data were flagged. \citet{Pe13} flux density scale was used. For eighteen epoch of observations, the quality of data was not good and we could not make reasonably good images. We did not use data for these days. While using data for CD and D configuration, we used lower {\it uv} cut-off to avoid strong background extended radiation from the Galactic plane. We have made correction for the primary beam using {\tt AIPS} task {\tt PBCOR}.

Amplitude calibration was conducted in reference to 3C 286 and phase calibration was based on observations of the nearby source J2007+404 or J2015+371. 
We have not performed self calibration due to lack of any strong point source in the field. The integration time used for solving amplitude and phase during calibration was 10 sec. The images of different days have variable noise levels. Since the source is bit far from the field centre, the noise level near the source position is high. Also small on-source time for individual observation resulted relatively high noise level. The noise level near the source position varied between 0.08 mJy beam$^{-1}$ and 18.7 mJy beam$^{-1}$. The median value of noise was 1.8 mJy beam$^{-1}$.

\begin{figure*}
\vbox{
\centerline{
\psfig{figure=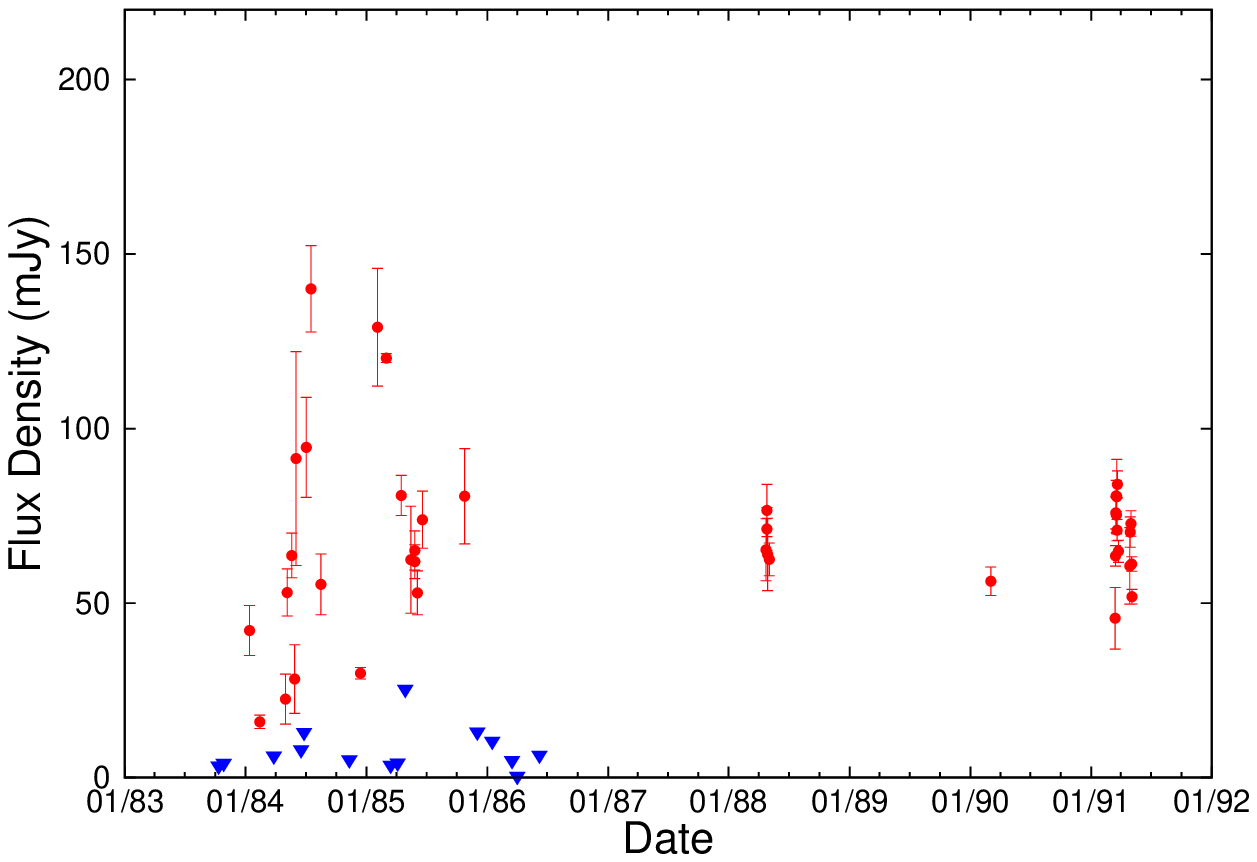,width=8.5cm}
\psfig{figure=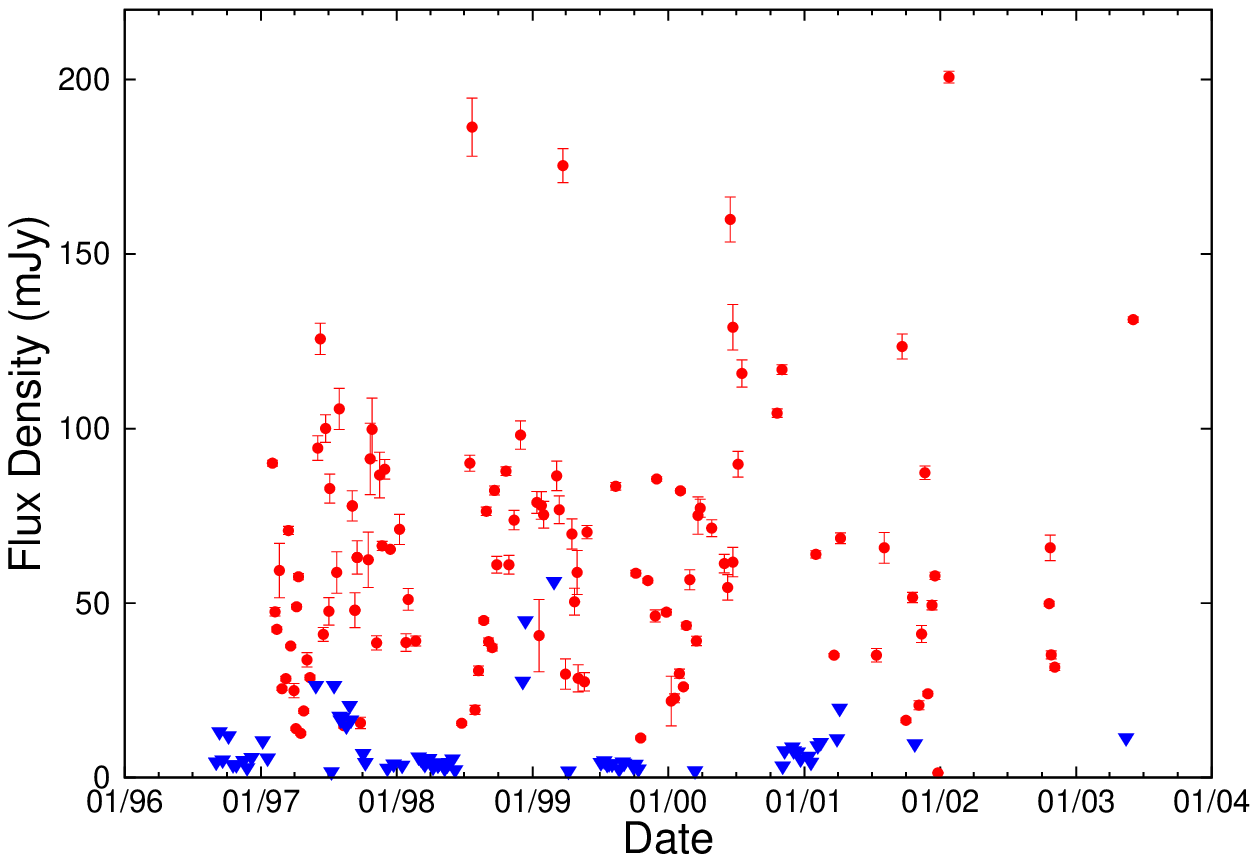,width=8.5cm}}}
\caption{Light curves of J195754+353513 at 1490 MHz. In left panel the observations between 1983 and 1991 are plotted and in the right panel the observations between 1996 and 2003 are plotted. The observations are done by The JVLA. Here the triangles show days of non-detection with $3\sigma$ upper limits.}
\label{light-curve}
\end{figure*}

Two models with Intermediate Frequencies (IFs) 1.3851 GHz (model 1) and 1.4649 GHz (model 2) were created combing data of 94 and 18 days respectively. Model with similar frequency configuration was subtracted with individual single epoch observations using the {\tt AIPS} task {\tt UVSUB} to look for variable sources. Apart from Cygnus X-1, we found that only one source J195754+353513 is present in many of the subtracted images with significantly fluctuating flux. The subtracted flux density for J195754+353513 was up to $\sim$120 mJy. The noise levels close to the location of J195754+353513 in model 1 and model 2 were 0.19 mJy beam$^{-1}$ and 0.46 mJy beam$^{-1}$ respectively.  

To be sure that the variation of J195754+353513 is not due to any kind of systematic effects or error in amplitude calibration, we measured flux density of another source NVSS J195823+345728 present in our field. The recorded flux density of the source in NVSS catalog is 52.3 $\pm$ 1.6 mJy \citep{Co98}. We found that during our study, the mean and median flux density of the source were 50.8 and 51.7 mJy respectively. The standard deviation in measurements of flux-densities of the source was 5.2 mJy which means the percentage of deviation was 9.9\%. This shows error in amplitude calibration play little role in large variation of J195754+353513, which will be discussed in more detail in coming sections.

\section{Results}
\subsection{Radio light curve of J195754+353513}
We have detected a transient radio source with co-ordinate 19h57m54.3s ($\pm$ 0.7s) +35$^{\circ}$34$^{\prime}$59.6$^{\prime \prime}$ ($\pm$ 0.6$^{\prime \prime})$ (J2000). This position is the average of result from fitting an elliptical Gaussian around the peak of the source, along with a background level to the source and the error is $1\sigma$ uncertainty in the fitting. The corresponding Galactic co-ordinates are $l=71.61^{\circ}$, $b=3.34^{\circ}$. After cross-correlation with NVSS sources, we found the transient source detected by us is NVSS J195754+353513. 

In the Figure \ref{transient-image}, examples of images of J195754+353513 are shown during observations in different JVLA configurations and time. The image of the source in B configuration on 7th November 1999 and 11th February 2000 with respective flux density 56.5 mJy and 26.0 mJy is shown in the upper left and upper right panels of the Figure. In the middle left, middle right and lower left panels, we have shown examples of detection of the source in A, C and D configurations of JVLA. In the lower right panel, an example of no detection is shown when JVLA was in C configuration. We have included the synthesized beam at the left-bottom corner of each panel. The observation details and different image parameters corresponding to Figure \ref{transient-image} is summarized in Table 2. 

The location of NVSS J195754+353513 is shown in all images of Figure \ref{transient-image}. The source is within extended emission region of J195754+353513. NVSS survey is carried out using D configuration of JVLA and the source is unresolved in NVSS with flux density 50.8 $\pm$ 1.6 mJy at 1400 MHz \citep{Co98}. It was not resolved during our study also when it was observed in D configuration.

The elongation of the source visible in the Figure \ref{transient-image} in north-south direction is most probably not due to intrinsic source structure but an effect of bandwidth smearing due to high channel width of 50 MHz. The direction of elongation of the source, as visible in Figure \ref{transient-image}, is towards the pointing center which is a signature of bandwidth smearing. In all the images, the source is consistently elongated in same direction, when detected.  

Light curves displaying variation of the source's flux densities between 1983 to 2003 is shown in Figure \ref{light-curve}. The flux density is calculated using {\tt AIPS} task {\tt JMFIT} using the results of fitting a Gaussian, along with a background level. The non-detections are reflected by the triangular points, which correspond to the $3\sigma$ upper limits. The error bars represent the rms noise levels of the images near the location of J195754+353513 and a 10\% uncertainty in the absolute calibration of each data set added in quadrature. There are only nineteen observations of the field between MJD 46612 (July, 1986) and MJD 50326 (August, 1996). The source was detected in 161 occasion and not detected in 83 occasions. The data was not good enough to make reasonably good images for 18 days. So, the source was successfully detected for 66.0\% days amongst the all observations with good data of the field. The source showed multiple short bursts and high variation. It varied from less than 0.3 mJy to 201 mJy. The median value of flux density was 38.9 mJy. On 25th January 2002 (MJD 52299) J195754+353513 reached maximum value within the observational period reported in this paper with 201 mJy flux density. The source showed signature of various flares and the peaks reached more than 150 mJy on 23rd July 1998 (186 mJy), 24th March 1999 (175 mJy) and 16th June 2000 (160 mJy). 

\begin{figure*}
\vbox{
\centerline{
\psfig{figure=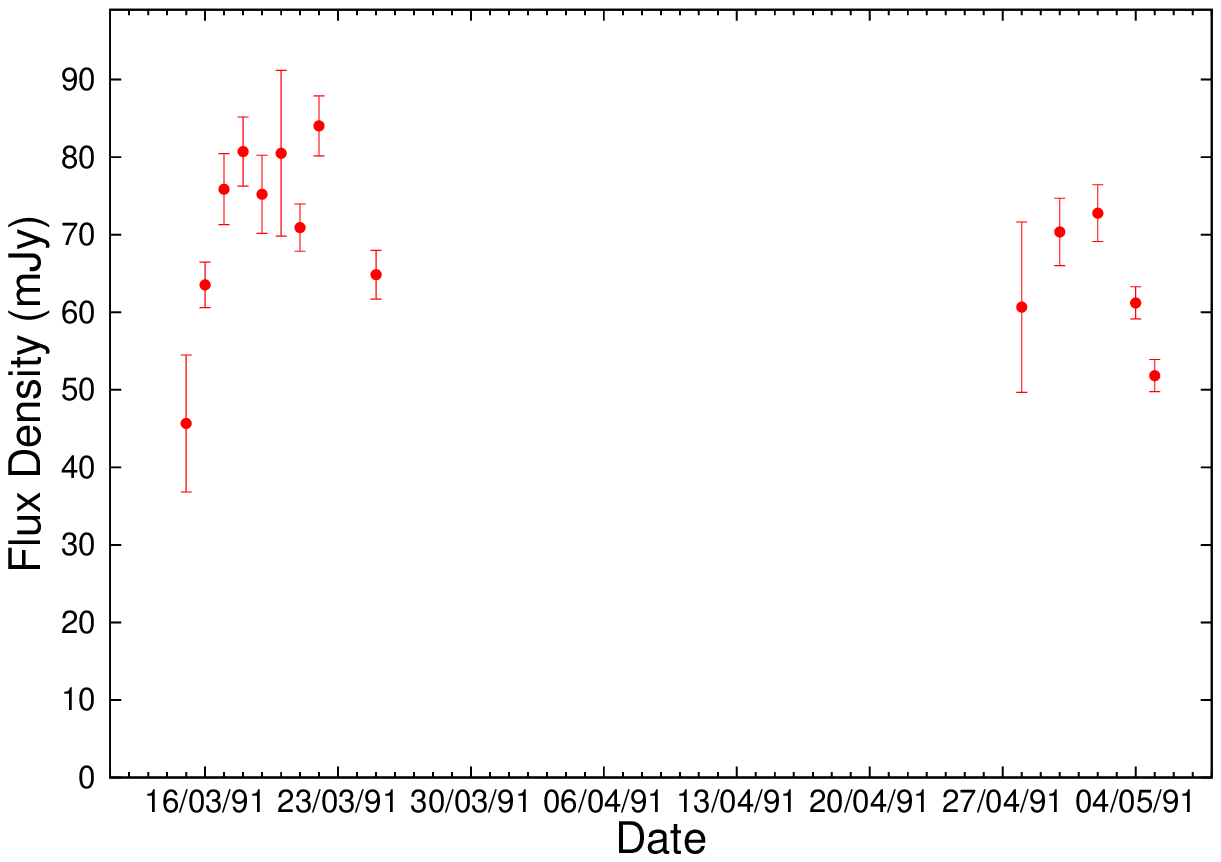,width=9cm,angle=0}
\psfig{figure=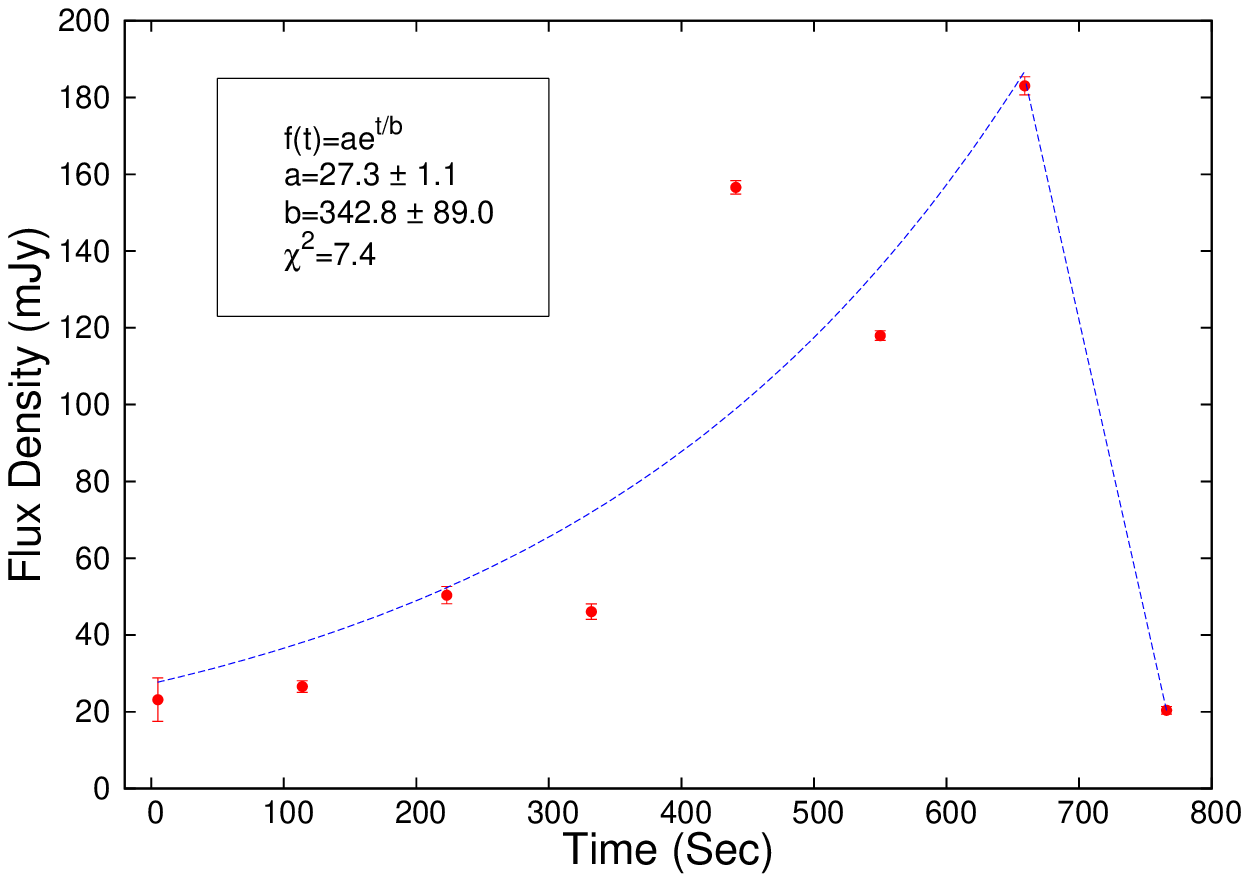,width=9cm,angle=0}}}
\caption{Left (Inter-day variation): Example of two flares of J195754+353513 observed at L band. These flares were observed in March and April-May 1991.  Antennas were in D configuration during these time. Right (Intra-day variation): An example of intra-day variation of J195754+353513 within a scan. The observation took place on 3rd February 2000. Antennas were in B configuration during this observation. Exponential fitting over the points in rising phase is also shown in the figure.}
\label{light-curve-zoom}
\end{figure*}

Due to gaps in the observations, we can study a complete flare only in a few cases. In Figure \ref{light-curve-zoom} we have shown two such flares in March and April-May of 1991 when the peak reached $\sim$85 mJy and $\sim$75 mJy respectively. Antennas were in D configuration during these times. Though there were not enough data points to precisely determine life-span and rise/decay rate of these flares, the approximate life span of these flares were in the range of 10--18 days. Both of these flares show sharp rise and relatively slow decay.

\begin{figure}
\vbox{
\centerline{
\psfig{figure=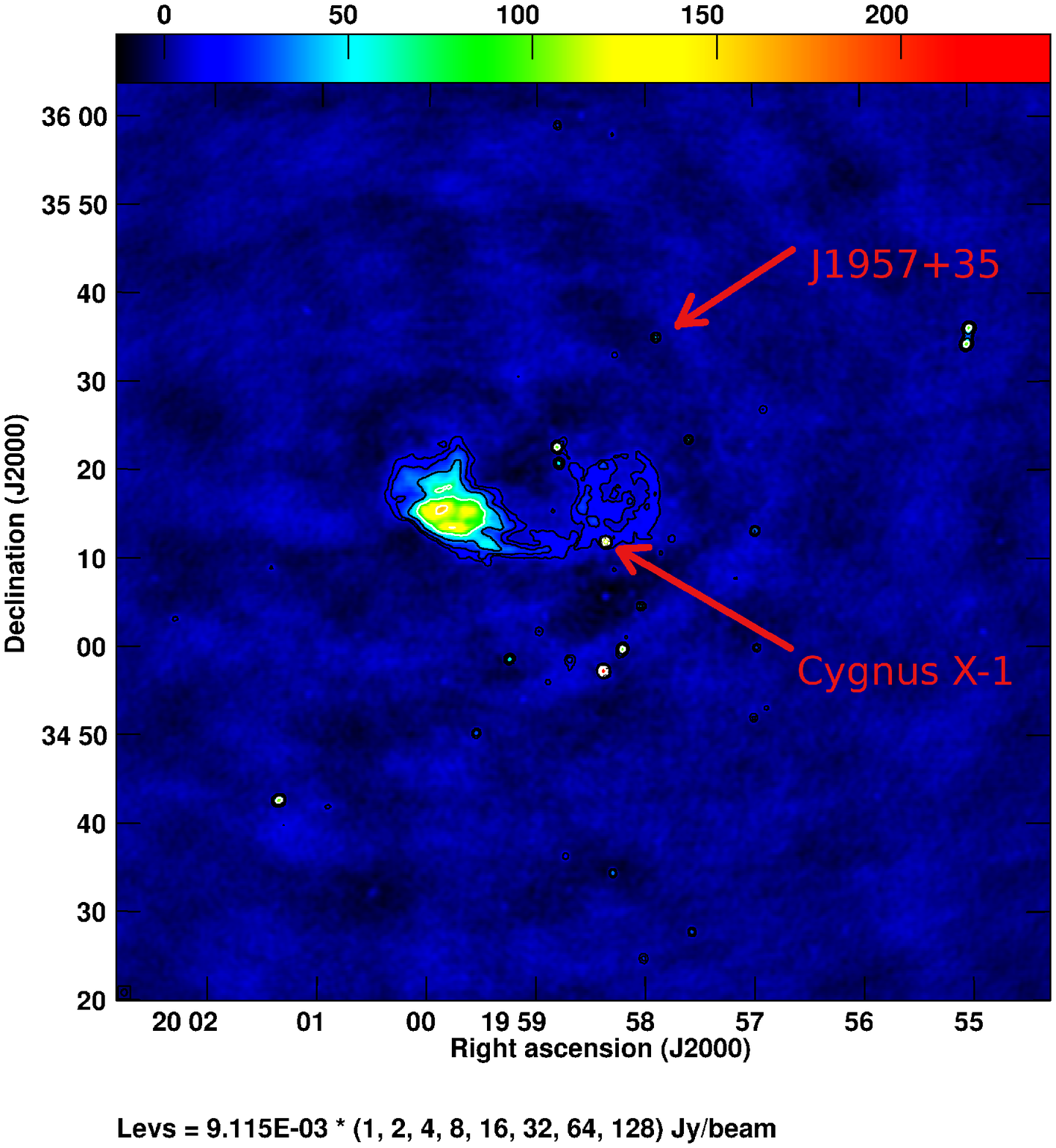,height=9cm,width=9cm}}}
\caption{1490 MHz JVLA image of the field of J195754+353513 combining 9 days of observations. All observations used in this image were conducted in 1991 when antennas were in D configuration. The RMS noise of the image is 0.17 mJy beam$^{-1}$. Cygnus X-1 and J195754+353513 are indicated by the arrow. The synthesized beam of the image is 43.24$^{\prime \prime}$ $\times$  41.71$^{\prime \prime}$.}
\label{image}
\end{figure}

In Figure \ref{image} we have shown the field of J195754+353513 combining 9 days of observations. All observations used in this Figure was made in 1991 when antennas were in D configuration. The RMS noise of the image was 0.17 mJy beam$^{-1}$ and the resolution of the image was 43.23$^{\prime \prime}\times$41.71$^{\prime \prime}$. We can clearly see J195754+353513 with 60 mJy flux density along with Cygnus X-1 and many other background sources. 

We looked for possible periodicity in J195754+353513 as was found in some other radio transients. The light-curve data was systematically folded with different period to search for signature of any periodicity in the emission. No periodicity was found in the source light curve.

There are no available JVLA archival data of the source in any other band except L band. Thus a study of the spectral index of the source was not possible using JVLA data. There is a source in 325 MHz WENSS (Westerbork Northern Sky Survey; \citet{Re97}) catalog which is 5.11 arcsec far from the NVSS position of the source. Since the positional uncertainty in WENSS for faint sources are $\sim$ 5 arcsec, this is most probably the same source. The source has a flat spectral index of $\alpha=-0.19$ (assuming $S_\nu \propto \nu^{\alpha}$) between NVSS and WENSS measurements \citep{Ma14b}. Since the source is highly variable and there is gap between measurements between NVSS and WENSS (observations of WENSS took place between 1991--1993 and observations of NVSS took place between 1993--1996), the flat spectral index measured between NVSS and WENSS is misleading.   

\subsection{Intra-day flux density variation}
\begin{table*}
\caption{\bf Intra-day Flux Density variation of J195754+353513}
\centering
\begin{tabular}{c c c c c c}
\hline
Date  &        1st Scan          &    1st Scan    & 2nd Scan   & 2nd Scan &Time gap\\
(UT)  &        Time Span         & Flux Density   & Time Span  & Flux Density& between scans        \\
      &        (min)          & (mJy)          & (min)   & (mJy)    & (min)\\
\hline
03/07/84 & 6.5 & 50.9 $\pm$ 8.4 & 6.5 & 216 $\pm$ 7.9 & 52.5\\
04/03/90 & 29.16&194 $\pm$ 0.7  & 29 & 31.4 $\pm$ 0.7&204.5\\
16/03/91 & 19.0  & 68.1 $\pm$ 2.4 & 13.5& 55.4 $\pm$ 2.1&5.5\\
25/03/91 & 15.5& 88.1 $\pm$ 3.6 & 14.5& 37.9 $\pm$ 2.6&5.5\\
16/03/97 & 0.92& 89.8 $\pm$ 1.2  & 0.92& 127 $\pm$ 2.8  &0$^{b}$\\
25/01/99 & 1.33& 111 $\pm$ 5.0  & 1.33& 48.4 $\pm$ 3.7  &0$^{b}$\\
24/10/02 & 2.83& 37.6 $\pm$ 2.6 & 3.17& 76.7 $\pm$ 2.8&57.5\\
\hline
\multicolumn{6}{l}{$^{b}$Note: Zero difference between two time-spans mean we are looking for intra-scan variations.}
\end{tabular}
\end{table*}

Based on previous studies of the bursting transient GCRT J1745--3009 \citep{Hy05, Hy07, Ro10} and GCRT J1742--3001 \citep{Hy09}, we searched for flux density variations of J195754+353513 between different scans of the observations of same day. We also imaged each minute separately, when the source is detected, to look for minute-scale variation of J195754+353513. Since the flux density of the source was not adequate to image it all the time with smaller time scale, we could make a study of the intra-day variation only for limited amount of time.

Significant scan to scan and intra-scan flux density variation is detected when it was possible to image separate scans. In the right panel of Figure \ref{light-curve-zoom} we have plotted an example of intra-day variation of J195754+353513 within a scan. The observation took place on 3rd February 2000 (MJD 51577). Antennas were in B configuration during this observation. The source rose from 20 to 180 mJy within 700 seconds and then came back to $\sim$20 mJy level within $\sim$100 seconds. The rise time constant, resulting from the exponential fit over the points in rising phase, is $\tau=342.8 \pm 89.0$ seconds. 
A power-law fit ($S_\nu \propto t^\beta$) over points in rising phase of the flare yields $\beta = 1.5 \pm 0.8$.

In Table 2, we have summarized intra-day flux density variations for different days. The error in flux density mentioned in the table is RMS noise near the location of the source. For five different days we found more than 20\% difference in flux densities between two successive scans of observations with time difference 5.5 minute to 204.5 minute. On 4th March 1990 (MJD 47954), the flux decayed from 194 mJy to 31.4 mJy with just 204.5 minute separation and on 3rd July 1984 (MJD 45884), the flux density rose from 50.9 mJy to 216 mJy when the gap between two successive scans were just 5.5 minutes. For two days, we found significant variation within a scan.  

The inter and intra scan variation, often more than twice in flux density, suggests the radio emission from J195754+353513 consists of many small bursts.

\subsection{Circularly polarized emission}

We looked for the presence of circularly polarized emission from J195754+353513 as was found for the case of GCRT J1745--3009 \citep{Ro10}. The measurement of Stokes $V$ polarization on 19th March 1991 (MJD 48334) was 12.6 mJy with RMS noise of 1.19 mJy beam$^{-1}$ and the measurement of Stokes $V$ polarization on 2nd May 1991 (MJD 48378) was 11.5 mJy with RMS noise of 1.36 mJy beam$^{-1}$.  
The corresponding value of $V/I$ on 19th March and 2nd May 1991 were $0.25$ and $0.24$ respectively with 3\% uncertainty in error. Relatively weak Stokes $V$ detection was done on 1st December 1999 (MJD 51513) with flux density 8.3 mJy and RMS noise 0.6 mJy beam$^{-1}$. The corresponding $V/I$ was $0.15$.
More than $5\sigma$ Stokes $V$ was not detected from data of any other days.

\subsection{Optical/IR and X-ray counterpart of the source}
No known sources are reported to emit X-ray emission from the nearby position of J195754+353513 making it impossible to detect radio emission from follow-up observation of X-ray emission at the time of flare. 

We have searched for optical/infra-red counterpart of J195754+353513. There are two sources within 10 arcsec from the source location of J195754+353513 in the Two Micron All-Sky Survey (2MASS) point source catalog \citep{Cu03, Sk06}. The nearest source in 2MASS catalog is J19575420+3535152 whose location is 1.82 arcsec away from J195754+353513 and may be the near infra-red counterpart of J195754+353513. The other source in 2MASS catalog within 10 arcsec from J195754+353513 is J19575378+3535221 whose position is 9.37 arcsec away from J195754+353513. The brightnesses of J19575420+3535152 in {\it J}, {\it H} and {\it K} bands are 15.45, 14.90 and 14.55 mags respectively.

\section{Discussion}

\subsection{The nature of the source J195754+353513}

We have searched the environment of J195754+353513 for associated discrete sources or extended structures. The source is not close to any known supernova remnant. We did not find trace of any extended structure close to the source, either. 

We have found one 2MASS source within 1$\sigma$ positional error of J195754+353513 and it is possible that the radio emission of J195754+353513 arises from activities in a foreground flaring star. Many flaring stars are known to exhibit activity in both radio and X-ray wavelengths such as the giant outburst from a young stellar object reported by \citet{Bo03}, but the detection of radio flares having no apparent associated X-ray emission is not uncommon. For example, radio flares from UV Ceti stars with seconds to minutes time-scale were detected at low frequencies by \citet{Sp74} and \citet{Ka77} with YZ Canis Minoris where no corresponding X-ray emission was found. At higher frequencies (4.9 and 8.4 GHz), \citet{Os05} reported short duration radio flares from the dMe flare star EV Lacertae which was not clearly related to the star's X-ray flares. The radio flares reported in these stars range from a few milli Jansky to a few tens of milli Jansky, with rise and decay times of $\sim1$ min and $\sim1$ hr respectively. We have also found small flares from J195754+353513 of $\sim700$ s duration as featured in Figure \ref{light-curve-zoom}.

\citet{Ri03} presented results on five years of continuous monitoring of radio flares of Algol-type and RS CVn systems; many of the flares reached hundreds of milli Jansky with a few days to a month duration. Numerous short bursts within each flare were also visible. Strong periodic activities are also found in these systems where the shortest periodicity was found in $\beta$ Per with 48.9$\pm$1.7 days. Though we could not detect any periodicity in emission from J195754+353513, the source could be RS CVn system due to similarity in time scale of flaring episodes. 

Pulsars can produce highly circularly polarized emission in single pulse (eg. \citet{Ka91}). There are some pulsars which show non-periodic flaring events like Crab pulsar (e.g. \citet{Bu12}). No known radio pulsar is reported from the nearby location of J195754+353513. Also we have not seen any sign of nearby supernova remnant or nebula. Though majority of pulsars are associated with supernova remnants \citep{Fr94}, some belief that this association are by chance and actually false \citep{Ga95a, Ga95b, Ga95c}. One should look for possible pulsar emission in the location of J195754+353513. 

Variations in the light curve of J195754+353513 could also be due to some kind of extreme scattering of the incident radiation in the interstellar medium of our Galaxy (e.g. \citet{Ri90, De02}). Variation up to few day time-scale can occur due to interstellar scintillation in GHz frequencies \citep{Pe00}. Since J195754+353513 showed variation of different time-scales, from minutes to months, even if interstellar scattering play some role, it is unlikely that all this variations are due to scattering effect. 

For the most of black-hole X-ray binaries, a universal correlation between radio and X-ray luminosities has been reported \citep{Co03, Ga03}. Assuming the relation given in \citet{Ga03}, the $\sim 200$ mJy peak should have corresponding X-ray flux of $\sim 0.25$ Crab in its hard state. Such a strong flare in X-ray would not go un-noticed by all sky X-ray monitors, given that the source is in flaring stage quite regularly. Though some of the exceptions are found for X-ray binaries which do not follow this universal correlation, all the examples have lower radio luminosity than the universal value but none of them have significant
lower X-ray luminosity. So, it is unlikely that the system is a black-hole X-ray binary system. 

On the other hand, if we consider a two component accretion flow (TCAF, \citet{Ch95}) where the Keplerian disk is flanked by a transonic flow in hard states when jets are stronger, it is well known that the Keplerian disk does not have to penetrate distances close to the black hole and the X-ray could be faint. We postulate that the disk with very low accretion rate could be located at a large distance with inner edge $\sim 20000$ Schwarzschild radii or more so that the source is presently having a quiescence stage, as far as X-rays is concerned. If this is the case, occasional outbursts every few to few tens of years is possible and one could look for them. It is also possible that the disk is shrouded by copious winds from the companion, much like SS 433 \citep{Ma84, Cl84}, where disk X-ray is completely blocked. In fact, the circularly polarized emission is an indication of aligned magnetic fields. If there is an outflow with a small inclination angle with the line of sight, a polarization fraction similar to those reported in Section 3.3 can be produced from its radio emission. However, in that case, a profusely mass-losing star star would be expected in the vicinity. As discussed in Section 3.4, it is possible that 2MASS J19575420+3535152 is the IR counterpart. The time-scale of rising and fading for every micro-flare would be expected to be of much shorter duration for a compact binary.

\subsection{Comparison with Galactic Center Transients}

There are many similarities between the temporal evolution of J195754+353513 to that of GCRT J1742--3001 and a transient radio source detected close to Galactic Center region (GCT). While GCRT J1742--3001 was detected as part of transient search program near Galactic centre region in March 2006 to May 2007 at 235 MHz \citep{Hy09}, GCT was detected in monitoring observations of Sgr A* from December 1990 to September 1991 at different radio wavelengths from 1.3 to 22 cm \citep{Zh92}. There are also some similarity with another transient GCRT J1745--3009 \citep{Hy05, Hy07, Ro10} located close to Galactic centre at 330 MHz. The flare of both GCRT J1742--3001 and GCT took about a month to rise while it decayed in about three months. For J195754+353513 we could not catch total rising and fading profile of many flares due to inadequate sampling rate and data gap, typical rising time of flares were $\sim$5 days and typical fading time was $\sim$10 days. Though the active time of individual flares of J195754+353513 were less than GCRT J1742--3001 and GCT, J195754+353513 exhibited higher frequency of such flares unlike other two sources mentioned above. GCRT J1742--3001 also showed fewer small bursts before the main flare and the GCT showed the presence of a significantly intense secondary burst in the 18--22 cm observations about six months after the primary burst. Though GCRT J1745--3009 was detected only three times, it showed intra-day variability with time-scale of hundreds of second as J195754+353513.   

GCRT J1742--3001 peak flux density was $\sim$100 mJy in 235 MHz and peak flux density of GCT was $\sim$1 Jy in the wavelength range 18--22 cm ($\sim$1.5 GHz). The peak flux density of individual flares of J195754+353513 varied between 30 to 200 mJy. Since GCRT J1742--3001 had a very steep spectrum ($\alpha \lesssim -2$, $S_\nu \propto \nu^\alpha$), its peak flux density should be fainter than J195754+353513 in 1400 MHz. Assuming $\alpha=-2$, we calculate peak flux density of GCRT J1742--3001 at 1400 MHz to be $\lesssim$21 mJy. The peak flux-densities for three detections of GCRT J1745--3009 were $\sim1$ Jy, $\sim0.5$ Jy and $\sim0.06$ Jy. The peak intensity may indicate strength of magnetic fields.

J195754+353513 showed presence of variable circular polarization as GCRT J1745--3009 \citep{Ro10}. Circular polarization was not reported for the case of GCRT J1742--3001 and GCT.

No X-ray counter-parts were confirmed for the case of GCT, GCRT J1742--3001 and GCRT J1745--3009. J195754+353513 also has no known X-ray counter part. There is some suggestion that GCT may be associated with X-ray binary \citep{Zh92}, which is yet to be established. 

\section{Conclusions}
We found transient nature of a radio source J195754+353513, approximately 24 arcminutes far from Galactic micro-quasar Cygnus X-1. The source showed high variability with different time-scales. J195754+353513 showed evidence of variable circular polarized emission. The source has no known X-ray counter part of the system. 2MASS J19575420+3535152 may be an NIR counterpart of the source. The nature of the source is still unknown. It is not unlikely that the system is a black-hole X-ray binary with the disk covered by significant amount of matter from the companion. On the other hand, it is also possible that the emission is coming from a foreground flare star. Clearly more multi-frequency monitoring is required to come to a definite conclusion.

\section*{Acknowledgments}
We acknowledge the anonymous referee whose detail and productive suggestions helped to improve the manuscript significantly. DP and SP acknowledge support of MOES fund to carry out this project. We have used data from JVLA which is run by The National Radio Astronomy Observatory (NRAO). NRAO is a facility of the National Science Foundation operated under cooperative agreement by Associated Universities, Inc.

\bsp

\label{lastpage}

\end{document}